\documentstyle[12pt]{article} 
\begin{document}

\begin{center}
{\large \bf     Self-Organized Criticality in  the Model\\
               of Biological Evolution Describing Interaction \\
\vskip2mm
               of "Coenophilous" and " Coenophobous" Species}
\vskip2cm

{\bf Oleg V.Kovalev }   \\
Zoological institute of Russian Academy of Sciences,\\
Universitetskaya Embankment 1, St.-Petersburg, 199034 Russia,\\
{\it Fax:} (7)(812)2182941, {\it E-mail:} kov@zisp.spb.su
\bigskip

{\bf Yuri M. Pis'mak
\footnote{ Supported in part by EC-Russia Collaboration Contract
ESPRIT P9282 ACTCS},
Vladimir V. Vechernin} \\
Department of Theoretical Physics, Institute of Physics,\\
State University of St.Petersburg, Ul'yanovskaya 1, Petrodvorez,\\
St.Petersburg, 198094, Russia\\
{\it Fax:} (7)(812)4287240, {\it E-mail:} pismak@niif.spb.su,
vechernin@niif.spb.su

\end{center}

\bigskip

The modification of the model of P.Bak and K.Sneppen of
the self-organized
 biological evolution is proposed on the basis of a formalization
 of the scheme of the biosphere evolution suggested by
 O.V.Kovalev. This scheme is regarded as one approximating
 the realistic model of the ecosystem evolution. The fundamental
 difference
 between "coenophilous" species and "coenophobous"
 ones in respect to their reaction on the external environment
 is represented.
 The dynamics of the modified model as
 well as that of the model of P.Bak and K. Sneppen possesses
 the most important features of selforganized criticality:
 the avalanche-like processes and the punctuated equilibrium.
 The results obtained by using the numerical
experiment for the study of these phenomena are presented.

\newpage

{\it Introduction} - Recently, the considerable attention has
been focussed on the studies of the phenomena of the self-organized
criticality (SOC) in dynamical systems. The main characteristic
feature of this type of the critical dynamics is its occurrence
without a special fine tuning of the system parameters.  Since only
the most fundamental properties of interactions manifest itself in
a critical dynamics there is a
good reason to believe that as the SOC
phenomena are the same in a number of real situations they could be
described in the framework of the universal theoretical models. The
 well studied model of such a kind is the "sand pile" model [1]. A
very interesting line of the SOC investigations is arisen on the
basis of the model of self-organized biological evolution suggested
by P.Bak and K.Sneppen [2].

This model describes the dynamics of the ecosystem of interacting species
governed by the processes of mutation and natural selection.
The SOC type in the Bak-Sneppen model (BSM) has the main specific
features of real biological evolution considered in the framework
of the Gould and Eldredge "punctuated equilibrium" conception [3].
The "punctuated equilibrium" manifests itself in the BSM critical
dynamics as alternating quasistable states and avalanche-like
processes of the balance disturbance. This model makes it possible
to explain nonuniform race and uneven character of the biological
evolution. The scale invariance of extinction events exposed with
help of analysis of paleontological data is in a good agreement
with criticality of BSM dynamics.

The interactions of species, mutations and a natural selection are
taken into account in the BSM. However, the factors of external
environment are not represented in an explicit form and the specific
of their influence on the trend of biological evolution is ignored.
In this paper we propose the modification of the BSM involving
consideration of the environments effects.

It is constructed
as formal mathematical modelling of important mechanisms of evolution
processes on the basis of "the model of evolution of biosphere"
proposed recently [4].

{\it Simple model of environment influence on the self-organization of
biological evolution}-
In the BSM framework, the species as an element of the ecosystem
has only one characteristic. It is its
average probability for survival in the population
called barrier, as it
is interpreted also as mutation stability.
The state of the ecosystem of $N$ species is considered to
be given at the time $t$ if the barriers $b_i(t)$,$i= 1,2,...,N $
are defined for all its species. The barrier $b_i(t)$ of the
$i$-th species is the function of a discrete time $t$ with values
belonging to the interval $[0,1]$: $0\leq b_i(t)\leq 1$.
The BSM dynamics
is formulated as follows. Initially, each barrier is set to a
randomly chosen value. At each time step the barrier
with minimal value of the "weakest" species and the barriers of
all species interacting with it (neighbors) are being replaced by new values.
In the random neighbor model (R) the neighbors of "weakest" species
are chosen at random. In the local or nearest neighbors model
(L) the interaction structure of the species is time
independent. Generally for its definition
the species are correlated with the
sites of the D dimensional lattice and the species corresponding
to nearest neighbors on the lattice are considered as the interacting
ones.  A new variable in our modification of the BSM is the time
dependent external environment factor (EEF) $f(t)$. We suggest that
$0\leq f(t)\leq 1$. The EEF influences the evolution processes
of the species. For to take it into account we introduce the characteristic
which will be called the type of reaction on the influence of
the environment (TRIE).
There are two alternative cases. The barrier transformation
of the species is determined only by the EEF and not influenced by
barriers of other species or the barriers of other species are also
essential for this process.  By definition, the first TRIE of the
species is caenophobous (Cpb) and the second TRIE will be called
caenophilous (Cpl).

Thus in the proposed BSM modification the state
of the species is given by its TRIE and its barrier.
We define the controlled
by EEF dynamics of this community of the species in
following way.  The initial state is to be chosen at random.  The EEF
is to be given at every time step.  At each time step, the state of
the Cpb species remains unchanged if the EEF is less then its
barrier.  If the EEF is greater then the Cpb
barrier then we have to assign the new random barrier value for this
species and to change its TRIE from the Cpb to the Cpl, if this new
barrier value will still less then the EEF.

As to the Cpl,
we have at first to define their interaction matrix
and to fix through it which species are interacting
with the given one. In principle the Cpl can also interact
with the Cpb but in this paper we restrict ourselves
by the case with the only Cpl mutual interaction.
In analogy with  the BSM we will consider
two types of interaction matrix: with the random neighbors
and with the nearest neighbors we will refer on these two
type of our model as MR and ML respectively.
At each time step we have to find the Cpl species
with the lowest barrier value, to assign the new
random barrier values for this species and
for the species which are interacting with it and
to change also the TRIE from the Cpl to the Cpb,
if this new species barrier value will less then the EEF.

If in accordance with our rules we are changing the TRIE
from the Cpb to Cpl we put this species at the random place
in the Cpl community both for MR and ML types of our model.
Respectively if we are changing the TRIE
from the Cpl to Cpb we extract this species from its place
in the Cpl community.
Correspondingly we have to change the interaction
matrix in these two cases.
Note that in the case of ML model this alters the definition of
the nearest neighbors in the vicinity of the species with
changing TRIE.

{\it Results of numerical experiments}-
If the function $f(t)$ representing the TRIE is constant, the system
arrives the stable state
(for the number of the time steps comparable with the number of
the species). All the species become Cpb with the uniformly
distributed on the interval $[f,1]$ barriers. If $f(t)=0$, there are
the stable Cpb subsystem of the species having initial barriers and
the Cpl subsystem evolving as the corresponding (random or linear)
BSM. For $f=1$ the dynamics of the Cpl subsystem is as follows:
on each time step the equal number of species leaves the subsystem
and is introduced in it.  As the species are introduced on the random
sites of the Cpl subsystem,  for $f=1$ the dynamical
behavior of the MR and ML systems is practically the same and
coincides with the case R of the BSM.

The dynamics with  the
stochastic EEF was investigated for the $f(t)$ uniformly
distributed on the interval $[0,1]$. It has the following characteristic
features. After starting avalanche-like processes the greatest part
of the species became Cpl. Then the number of Cpb species obtains the
stable tendency to increase. The common "punctuated equilibrium"-like
processes take place in the system. It can be considered as a
quasistable Cpl ecosystem (community) interacting by exchanging of
the species with the Cpb environment. Only the "weak" species are
involving in the exchange process. They are mainly "young", i.e. not
long ago leaving the Cpl ecosystem. The "strong" caenofobous species
having high barriers are "in general situation" not introduced in
the caenophilous ecosystem (it could be called the
"non-caenophilisation" trend of the evolution).  The most of the
"strong" Cpb species are "old", i.e. they evolve long ago in the
"non-caenophilisation" kind.  The "exchange interaction" influences
   the "punctuated equilibrium" dynamics of the Cpl ecosystem
(community), and results in its disintegration and in the increase of
the number of the "old", "strong" Cpb species. The disintegration
processes are $10^2--10^3$ times slower than the processes of the
  reaching of "the punctuated equilibrium".

   On the Fig.1 and  Fig.2 the results of numerical simulations are
shown for the distributions of the barriers ("right" curves) and  the
minimal barrier ("link" curves). In the Fig.1 (Fig.2), ones are
presented for the ML(MR)-model for the cases: $f=0$ (stars),
$f=1$ (crosses), the stochastic $f$ (squares). As it was mentioned
above, the "$f=0$"-curves present the common BSM distributions
respectively for the cases L(R).
The "$f=1$"-curves of ML and MR models coincide practically
 with each other and with the R case of the
corresponding BSM distributions.
For the stochastic case, the very similar curves have been obtained
for the MR and ML models, i.e. for the random EEF the behavior
of the systems is practically universal and independent on detail
properties of the interactions. It is interesting that the
curve for the "stochastic"
minimal barrier distribution have an clearly expressed
maximum at $b\neq0$, whereas the distribution functions for the
minimal barrier are monophonically decreasing in the BSM (the cases
$f=0$ and $f=1$).

In the Fig.3 the distribution for the space correlations of the
minimal barrier
in the Cpl community of the ML-model is shown
for $f=0$ (stars), $f=1$ (crosses) and the stochastic $f$ (squares).
For $f=0$ it coincides with the space correlations distribution
for the L case of the BSM.
For $f=1$ the distribution is uniform as it must be for
 the R case of the BSM.
For the random $f$ one sees that there are only
"short-range" space correlation in the "punctuated equilibrium".

To obtain the distribution function of the Cpb species one needs the
longer time period
of numerical simulations. The point is that at the first stages
of dynamical processes their portion in the whole community is very small.
Our numerical experiments show that for the sufficiently large time
$t$ the barrier distribution $P(t,b)$ can be approximated  as
follows:
$$ P(t,b) = \frac{\alpha(t)}{(1-b)^{1-\alpha(t)}},
$$
where
$\alpha(t)>0$ is universal for the MR and ML models and becomes very
small for the large $t$. Thus the average value of the Cpb barriers
is very high, i.e. as it was mentioned above, the most
species of the Cpb community are "strong" and "old" and the exchange
interaction between Cpb and Cpl communities involves few number of
the "weak" and "young" species.

{\it Conclusion}-
Summing up, one can draw the following inferences. In our BSM
modification the type of the interspecies interaction
essentially manifest itself
only in the case, when the influence of the EEF is
not strong enough. The greater is the role of EEF in the evolution
processes the less role plays the interacting type in the dynamics
of the Cpl community. One can note the following peculiarities of
the behavior of this community in MR and ML models with respect to
original RBSM and LBSM models. The intensification of the EEF
influence on the evolution process decreases the average barrier
value in the ML model: for $f=0$ the average barrier is greatest, for
$f=1$ it is the lowest, for the stochastic case the average barrier
is intermediate. The situation is a very different for the MR-model:
the average barrier is lower for $f=1$ (maximal effect of the EEF on
dynamics) and $f=0$ (the EEF does not influence on dynamics) whereas
for the random $f$ (the intermediate effect) the average barrier is
greater, i.e. the random EEF increases the average survival
probability of the Cpl species in the system with random interaction.

In conclusion, we discuss the biological ideas which have stimulated
constructing of the considered model. The BSM was elaborated on the
basis of the Gould-Eldredge "punctuated equilibrium model", in which
the EEF role in ecosystem evolution was not considered. However,
there is the reason to believe that it is impossible to model
biological processes regardless the external factors. The community
evolution is controlled by the climate evolution and the climate
transformations are the integral reflection of several
geological, geophysical and cosmical processes [4]. Influenced by the
climate transformations (the EEF in our model) the succession
processes form in the nature the consequences of the community series
from the quasisteady to stable states. The importance of the
difference between the "caenophilous"- and "caenophobous"-like
species for the evolution process is in a good agreement with the
MacArthur's theory of "K"- and "r"-selection [5] and with the
Krasilov's "ecosystem theory of evolution" [6]. In the framework of
MacArthur`s conception the "r-selection" is dominating for the
fecundity and colonizing and the "K-selection" ensures the efficiency
and adaptiveness [5],[7].  In Krasilov`s approach [6] the model of
"coherent" and "non-coherent" trends of evolution is elaborated.  At
a period when the ecosystems decay (the non-coherent phase of
evolution) the "pioneer" ("caenophobous") species become the active
colonisators quickly capturing niches under conditions of the weak
concurrence.  In the "model of the evolution of biosphere" [4],
"caenophobous" species or "philogenetically advanced juvenil taxa"
have  in virtue of low caenotic intensity the advantage of the genom
structure:  activity of the mobile genetic elements, poliplody,
parthenogenesis et.c. This is why  in the processes of permanently
changing "external factor" of climate genesis, the key events of
evolution process are the interrelations between "caenophobous" and
"caenophilous" species.

As we have seen, there are the similar dynamics features  in our model.
At a period when the EEF are changing weakly $(f\approx const, f\neq
0, f\neq 1)$, the avalanche-like increase of the Cpb species number
happens. Correspondingly the number of Cpl species quickly decreases,
what can be interpreted as the decay of the Cpl ecosystem and capturing of
empty ecological niches by the Cpb species.  The repeating sharp EEF
changes result in the Cpl species prevalation over the Cpb ones due
to the domination of the coherent evolution processes under this EEF.
So it seems to us that the main dynamical properties of our model
reflect the specific features of the real biological evolution which
can not be described in the framework  of the more simple BSM.

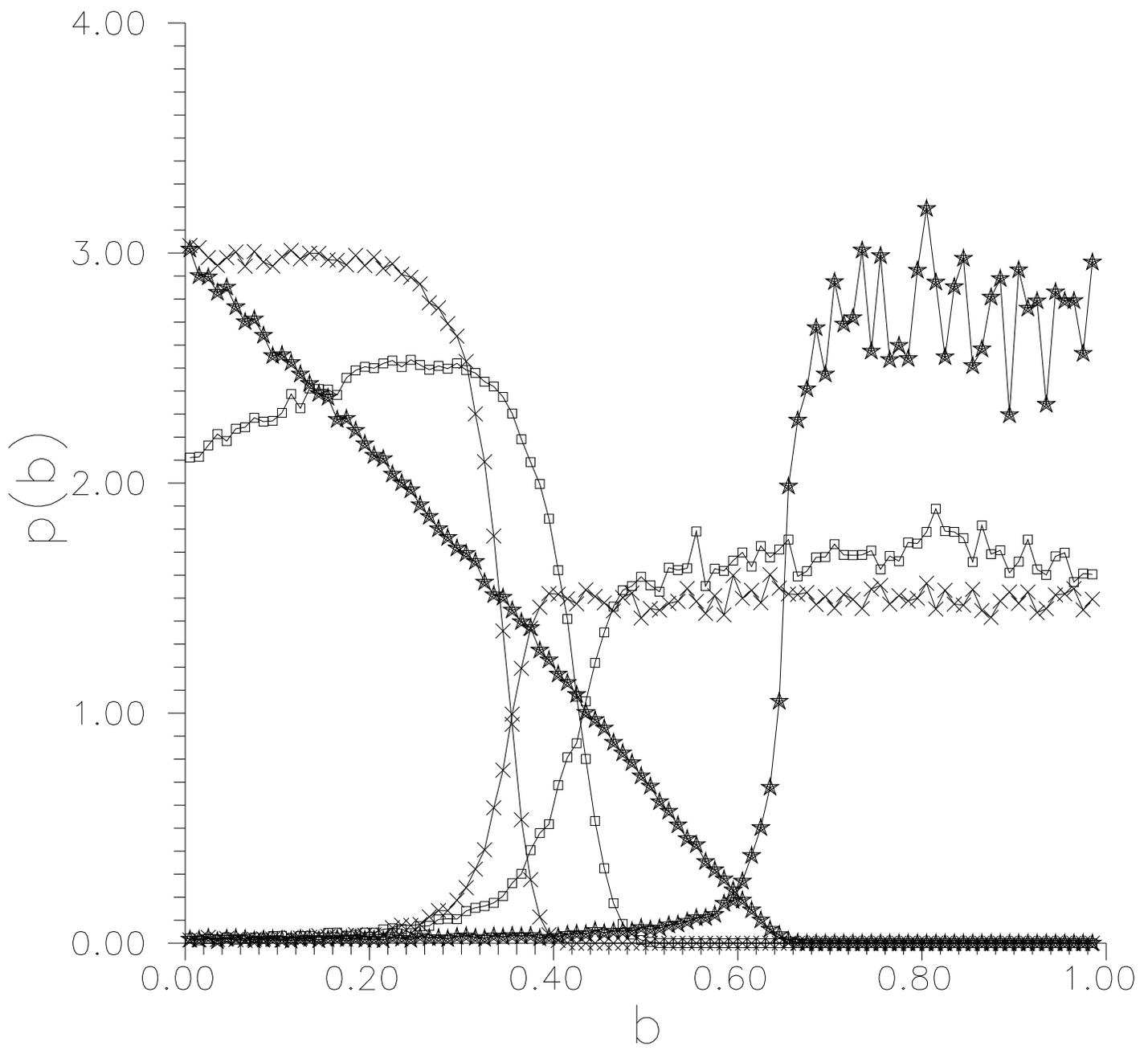
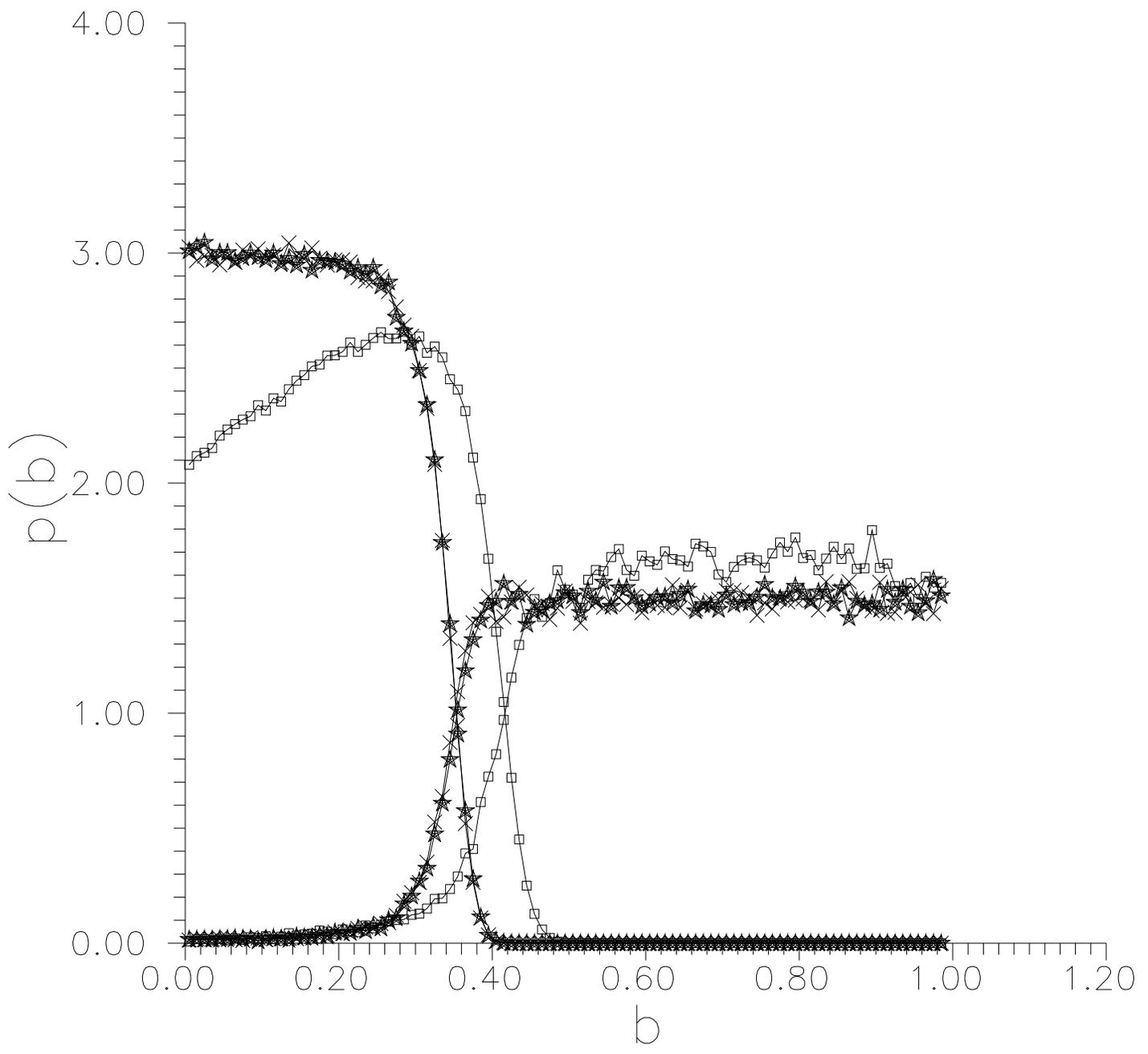
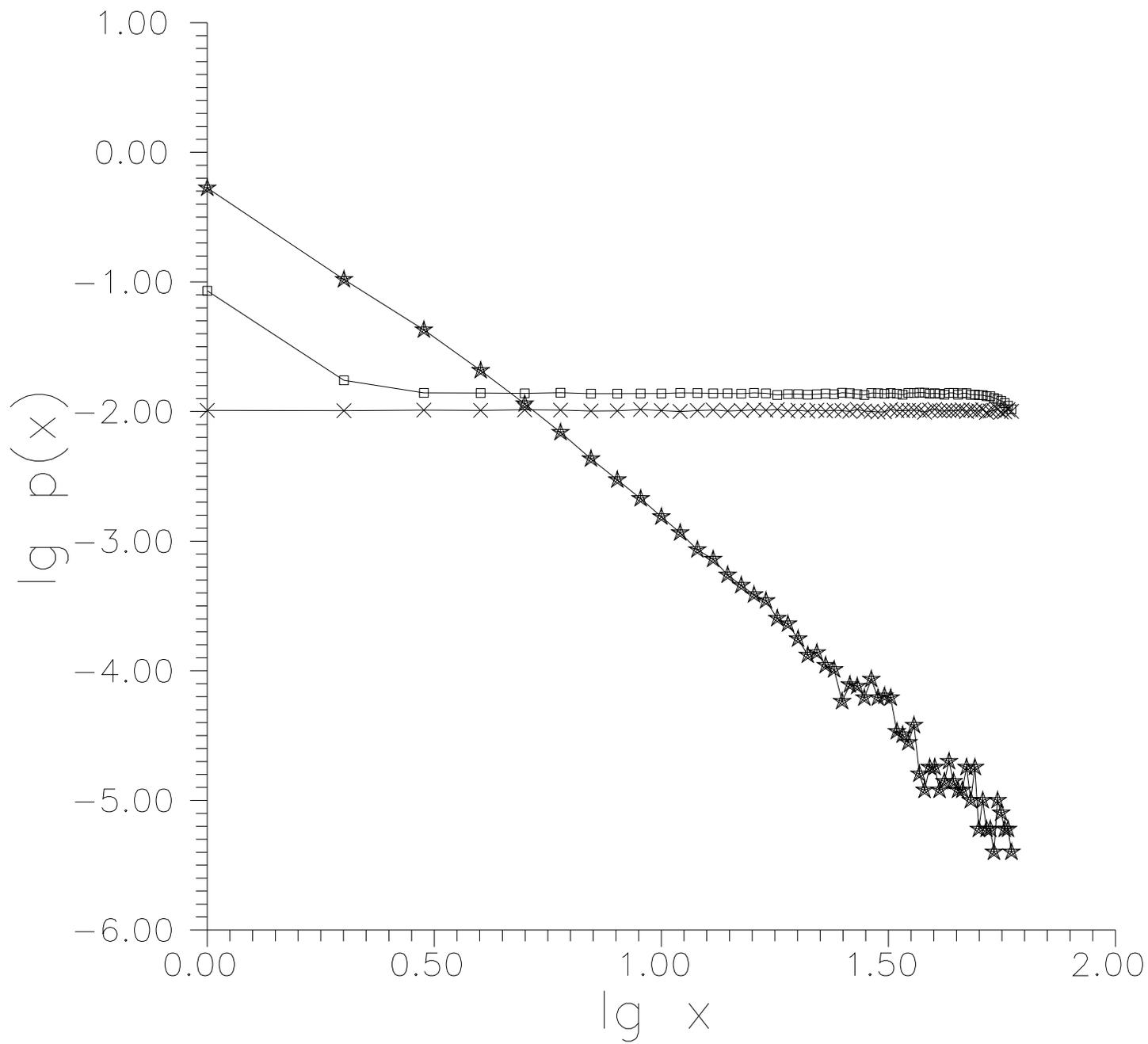
\end{document}